\documentclass[10pt,letterpaper]{article}
\usepackage[top=0.85in,left=2.75in,footskip=0.75in]{geometry}

\usepackage{amsmath,amssymb}
\usepackage{changepage}
\usepackage{textcomp,marvosym}
\usepackage{cite}
\usepackage{nameref}
\usepackage[right]{lineno}
\usepackage[nopatch=eqnum]{microtype}
\DisableLigatures[f]{encoding = *, family = * }
\usepackage[table]{xcolor}
\usepackage{array}
\usepackage{pifont}  % for check and cross marks

\usepackage{pifont}       % for check and cross marks
\usepackage{changepage}
\usepackage[hidelinks]{hyperref}

\newcolumntype{+}{!{\vrule width 2pt}}
\newlength\savedwidth

\raggedright
\usepackage[aboveskip=1pt,labelfont=bf,labelsep=period,justification=raggedright,singlelinecheck=off]{caption}

\makeatletter
\renewcommand{\@biblabel}[1]{\quad#1.}
\makeatother

\usepackage{lastpage,fancyhdr,graphicx}
\usepackage{epstopdf}
\fancyheadoffset[L]{2.25in}
\fancyfootoffset[L]{2.25in}
\begin{document}
\vspace*{0.2in}

\begin{flushleft}
{\Large
\textbf{Evaluating Global Measures of Network Centralization: Axiomatic and Numerical Assessments}
}
\newline
\\
Majid Saberi\textsuperscript{1*},
Samin Aref\textsuperscript{2*}
\\
\bigskip
\textbf{1} Headache and Orofacial Pain Effort (H.O.P.E.), Biologic and Materials Science \& Prosthodontics, University of Michigan School of Dentistry, Ann Arbor, MI, USA
\\
\textbf{2} Department of Mechanical and Industrial Engineering, University of Toronto, Toronto, ON, Canada
\\
\bigskip
* \texttt{majidsa@umich.edu}, \texttt{s.aref@utoronto.ca}
\end{flushleft}

\section*{Abstract}
Network centralization, driven by hub nodes, impacts communication efficiency, structural integration, and dynamic processes such as diffusion and synchronization. Although numerous centralization measures exist, a major challenge lies in determining measures that are both theoretically sound and empirically reliable across different network contexts. To resolve this challenge, we normalize 11 measures of network centralization and assess them systematically using an axiomatic framework and numerical simulations. Our axiomatic assessment tests each measure against the six postulates of centralization, ensuring consistency with minimal theoretical requirements. These include calibration at 0 and 1, invariance to isomorphism, and proper response to node degree saturation. In addition, our numerical assessment examines the behavior of normalized centralization measures over different random graphs (star, ring, complete) and their perturbed variants for which desirable trends and values of centralization are expected. Our results indicate major differences among the measures, despite their common aim of quantifying centralization. Together, our assessments point to the relative suitability of three measures: normalized betweenness centralization, normalized closeness centralization, and normalized degree centralization. Applying these three measures to real-world networks from diverse domains reveals meaningful variation in the organization of the networks with respect to hubs. Normalized betweenness centralization highlights path-based dominance; normalized closeness centralization reflects accessibility and efficiency of reach; and normalized degree centralization captures degree-based hub concentration. When used jointly, the three measures demonstrate the required sensitivity to varying levels of centralization and provide complementary aspects of network centralization that no single measure can offer alone. Our dual evaluation framework clarifies conceptual differences among existing measures and offers practical guidance for selecting reliable centralization metrics. 

\bigskip
\textbf{Keywords:} Complex networks; Network hubs; Hub; Centralization; Centrality; Hubness

%\linenumbers

\section*{Introduction}
Centralization is a fundamental structural property of complex networks, reflecting the extent to which one or a few nodes dominate in terms of centrality. Highly centralized networks tend to be organized around one or a few hubs: nodes with substantially higher degrees than the average node. This plays a critical role in shaping network structure and dynamics~\cite{bib1,bib2,bib3,bib4,bib5}. Such hubs can maintain efficient communication pathways, regulate system-level behavior, and influence diffusion processes including information spread, epidemic outbreaks, and synchronization dynamics~\cite{bib6,bib7}. Network hubs emerge in diverse contexts: influential users in social media~\cite{bib8,bib9}, essential proteins in protein–protein interaction networks~\cite{bib10,bib11}, hub airports in global transportation~\cite{bib12,bib13}, and integrative brain regions in neural systems~\cite{bib14,bib15}. 

While node-level centrality measures such as degree centrality, betweenness centrality, or closeness centrality are effective for identifying individual hubs~\cite{bib5}, they do not capture the global property of a network to have such dominant nodes. This network-level property, often referred to as centralization, is essential for understanding the extent to which the structure exhibits dominant hubs~\cite{bib1,bib16,bib17}. It is an open challenge to quantify centralization in a way that is theoretically sound, interpretable, and comparable across networks of varying size and density.

Several existing global centralization measures are based on centrality measures. Degree centralization~\cite{bib1}, the most widely used centralization metric, compares the node degrees to that of a maximally centralized network. Betweenness centralization~\cite{bib1} assess domination in terms of shortest-path control, while closeness centralization~\cite{bib18} relies on a notion of centralization related to geodesic proximity.

Other existing centralization measures include the degree (distribution) variance~\cite{bib19,bib20,bib21}, hub dominance~\cite{bib22}, the Gini coefficient~\cite{bib23,bib24}, hub formation tendency~\cite{bib15}, spectral centralization~\cite{bib25,bib26}, and entropy-based centralization~\cite{bib27}. Although these metrics have been used widely for measuring centralization, they differ in their mathematical foundations, normalizations, interpretability, and values. Some measures, such as degree variance or the Gini coefficient, provide statistical descriptions but lack explicit topological reference points. Other measures, such as hub dominance, are simple and interpretable but may not fully reflect nuanced patterns of centralization.

A star network has been described by numerous authors to represent the extreme centralization~\cite{bib1,bib16,bib28}. There are multiple existing measures that are aligned with such characterization of maximum centralization. However, without a formal set of requirements, any proposed metric could be labeled a centralization measure, making systematic comparison difficult. Palak and Nguyen have proposed a set of six axioms specifying minimal theoretical requirements for such measures, providing a principled framework for evaluation~\cite{bib16}. To the best of our knowledge, no other study has systematically assessed both the axiomatic validity and numerical behavior of centralization measures across different network types.

In this study, we conduct a comparative evaluation of 11 normalized centralization measures, including centrality-based metrics, statistical, and spectral measures. We take a two-pronged approach:
\begin{enumerate}
    \item \textbf{Axiomatic assessment:} We evaluate each measure against the six axioms of Palak and Nguyen~\cite{bib16}, identifying their theoretical strengths and weaknesses based on their alignments with the axioms.
    \item \textbf{Numerical assessment:} We analyze the behavior of the measures across synthetic networks for which a desirable behavior is expected. This involves testing measures for expected patterns on star graphs, ring graphs, complete graphs, and their perturbed variants.
\end{enumerate}

We integrate these results into a combined performance score, identifying the three most robust measures out of 11 measures. We then apply the three measures to a diverse set of real-world networks from neuroscience, biology, social systems, and ecology, illustrating their individual strengths and their complementarity. This unified framework clarifies the conceptual and practical differences among centralization measures, enabling researchers to make informed choices when analyzing network centralization in different contexts and disciplines.

\section*{Definitions of Centralization Measures}

Centralization measures quantify the presence and intensity of hubs in networks. It is customary that measures are normalized to the range $[0,1]$, where larger values indicate higher centralization. If an existing measure was not normalized already, we normalize it to ensure consistency and comparability across measures.

Let $G = (V, E)$ be a simple undirected graph with $|V| = n$ nodes and $|E| = m$ links and let $d_i$ denote the degree of node $i \in V$.

\textbf{Assortativity-Based Hubness (ABH)} can be defined for graphs with $n > 1$ and $m > 0$ as:
\[
ABH(G) = \frac{1 - r(G)}{2}
\]
where $r(G)$ is the degree assortativity coefficient of graph $G$~\cite{bib29}. In star graphs, $r(G) \to -1$ as $n$ increases, while in regular graphs $r(G) = 1$. This measure ranges from 0 (maximally assortative) to 1 (maximally disassortative). For $n < 2$ or $m = 0$, we define $ABH(G) = 0$.

\textbf{Eigenvector Centrality Dispersion (ECD)} is a normalized measure of inequality in eigenvector centrality among nodes~\cite{bib30}. For graphs with $n > 1$, it is defined as:
\[
ECD(G) = 
\frac{\sqrt{\sum_{i=1}^{n} \left(C_E(i) - \bar{C}_E\right)^2 / n}}
{\sqrt{(n-1)/n^2}}
\]
where $C_E(i)$ is the L2-normalized eigenvector centrality of node $i$ and $\bar{C}_E$ is its mean across all nodes. The denominator represents the maximum possible value of the numerator over graphs with $n$ nodes. The maximum value occurs in certain sparse, disconnected graphs with degree-1 nodes (pendant nodes). For $n < 2$, $ECD(G) = 0$.

\textbf{Normalized Betweenness Centralization (NBC)} for graphs with $n > 2$, measures domination of central nodes in terms of shortest-path control~\cite{bib18}:
\[
NBC(G) = 
\frac{\sum_{i=1}^{n} [C_B^* - C_B(i)]}{\max(\sum_{i=1}^{n}[C_B^* - C_B(i)])}
= \frac{\sum_{i=1}^{n} [C_B^* - C_B(i)]}{n - 1}
\]
where $C_B(i)$ is the normalized betweenness centrality of node $i$ and $C_B^*$ is its maximum value over all nodes of $G$. The denominator equals the maximum value of the numerator over all graphs with $n$ nodes, which is realized in star graphs. For graphs with $n < 3$, we define $NBC(G) = 0$.

\textbf{Normalized Closeness Centralization (NCC)} is defined for $n > 2$ based on the closeness centrality of nodes~\cite{bib1}:
\[
NCC(G) =
\frac{\sum_{i=1}^{n}[C_C^* - C_C(i)]}
{\max(\sum_{i=1}^{n}[C_C^* - C_C(i)])}
= \frac{(2n - 3)\sum_{i=1}^{n}[C_C^* - C_C(i)]}{(n - 1)(n - 2)}
\]
where $C_C(i)$ is the normalized closeness centrality of node $i$ and $C_C^*$ is its maximum over all nodes of $G$. The denominator equals the maximum value of the numerator over all graphs with $n$ nodes, realized in star graphs. For $n < 3$, we define $NCC(G) = 0$.

\textbf{Normalized Degree Centralization (NDC)}, also known as Freeman centralization or Palak-Wojtkiewicz centralization~\cite{bib1,bib31}, is defined for $n > 2$ as:
\[
NDC(G) =
\frac{\sum_{i=1}^{n}(d_{\max} - d_i)}
{\max(\sum_{i=1}^{n}(d_{\max} - d_i))}
= \frac{n(d_{\max} - \bar{d})}{(n - 1)(n - 2)}
\]
where $d_{\max}$ is the maximum node degree and $\bar{d}$ is the average node degree. The denominator corresponds to the maximum possible value of the numerator, realized in star graphs. For $n < 3$, $NDC(G) = 0$.

\textbf{Normalized Degree Entropy (NDE)} is defined for graphs with $n > 1$, based on the Shannon entropy of the degree distribution~\cite{bib32}:
\[
NDE(G) = -\frac{1}{\ln(n)} \sum_{k=0}^{n-1}\left(\frac{n_k}{n}\ln\frac{n_k}{n}\right)
\]
where $n_k$ is the number of nodes of degree $k$. The maximum occurs when all degrees are unique. For $n < 2$, $NDE(G) = 0$.

\textbf{Normalized Degree Variance (NDV)} is defined for graphs with $n > 2$~\cite{bib19,bib20}:
\[
NDV(G) = 
\frac{\sum_{i=1}^{n}(d_i - \bar{d})^2 / n}{DV_{\max}},
\]
\[
DV_{\max} = \max\left(\frac{(n-1)(n-2)^2}{n^2}, \frac{(2n^3 - 6n) - (4n - 6)^2}{n^2}\right).
\]
The denominator term $DV_{\max}$ is defined based on the observation that the maximum variance of degrees occurs in a star graph for $n < 7$ and in a two-hub graph for $n \ge 7$. For $n < 3$, $NDV(G) = 0$.

\textbf{Normalized Gini Coefficient (NGC)} is defined for graphs with $n > 2$~\cite{bib33}:
\[
NGC(G) = 
\frac{\sum_{i=1}^{n}\sum_{j=1}^{n}|d_i - d_j| / (2n^2 \bar{d})}{(n - 2)/n}
\]
The denominator equals the maximum value of the numerator over graphs with $n$ nodes, achieved in a one-edge graph with $n - 2$ isolated nodes. For $n < 3$, $NGC(G) = 0$.

\textbf{Normalized Hub Dominance (NHD)} is defined for $n > 1$~\cite{bib22}:
\[
NHD(G) = \frac{d_{\max}}{(n - 1)}
\]
For $n < 2$ or $m = 0$, $NHD(G) = 0$. While the NHD is particularly simple, it is often used in the context of quantifying hubness for a community~\cite{bib22}. In that case, the measure is based on dividing the highest degree among nodes of a community by $(n - 1)$ for the subgraph induced by that community. In such a usage, the measure may take values larger than 1.

\textbf{Normalized Hub Formation Tendency (NHT)} is defined for $n > 1$ and $m > 0$~\cite{bib15}:
\[
NHT(G) = 
\frac{\sum_{i=1}^{n} d_i^2 / \sum_{i=1}^{n} d_i}{(m + 1)/2}
\]
The denominator equals the maximum value of the numerator over graphs with $m$ edges, achieved in a star topology. For $n < 2$ or $m = 0$, $NHT(G) = 0$.

\textbf{Normalized Natural Connectivity (NNC)} is defined for graphs with $m > 0$~\cite{bib34}:
\[
NNC(G) = \frac{\lambda_{\max} - \bar{\lambda}(G)}{\lambda_{\max}}
\]
\[
\lambda_{\max} = \ln\left(\frac{e^{n-1} + (n - 1)e^{-1}}{n}\right), \quad 
\bar{\lambda}(G) = \ln\left(\frac{1}{n}\sum_{i=1}^{n} e^{\lambda_i}\right)
\]
where $\lambda_i$ are the adjacency matrix eigenvalues of graph $G$. For $m = 0$, we define $NNC(G) = 0$.

\section*{Axiomatic Framework for Centralization Measures}

Let $G = (V, E)$ be a simple undirected graph, with $|V| = n$ nodes and $|E| = m$ edges. $C(G)$ denotes the centralization value of graph $G$ based on an arbitrary centralization measure. A saturated node is one with degree $|V| - 1$, meaning it is connected to all other nodes. Saturating a node refers to adding edges to a non-saturated node so that it gets connected to all other nodes. The six postulates of Palak and Nguyen (2021)~\cite{bib16} can be expressed as follows:\\

\begin{itemize}
    \item 
P1a: If $|V| = 1$, then $C(G) = 0$. 
    \item 
P1b: If $G$ is complete, then $C(G) = 0$. 
    \item 
P1c: If $|E| = 0$, then $C(G) = 0$. 
    \item 
P2: If $G$ is a star graph, then $C(G) = 1$. 
    \item 
P3: If $G_1$ and $G_2$ are isomorphic, then $C(G_1) = C(G_2)$. 
    \item 
P4: If $G$ has no node of degree $|V| - 1$, then $C(G) < 1$. 
    \item 
P5: If $G$ has at least one saturated node and at least one non-saturated node, then saturating a node of $G$ to obtain graph $Y$ will not increase the centralization in the resulting graph $Y$, i.e., $C(G) \ge C(Y)$. 
    \item 
P6: If $G$ has no saturated node, then saturating a node of $G$ to obtain graph $Y$ will not decrease the centralization in the resulting graph $Y$, i.e., $C(G) \le C(Y)$.
\end{itemize}

These six postulates are intuitive and are justified in Palak and Nguyen (2021)~\cite{bib16}. The three parts of P1 define minimal centralization and calibrate it to the value 0. P2 defines maximal centralization to be required for a start graph, and calibrates it to the value 1. A star graph has been consistently identified as the graph with the highest centralization by several researchers~\cite{bib1,bib16,bib28}. P3 follows directly from the equivalence of degree distributions in isomorphic graphs. P4 specifies maximal centralization to require the presence of a node with the maximum degree. P5 describes a \textit{centralization decrease principle}: in a network that already has at least one saturated node, saturating another node cannot increase centralization because the existing saturated nodes become less unique and less dominant. P6 describes a \textit{centralization increase principle}: in a network with no saturated nodes, creating the first saturated node cannot decrease centralization because it makes that node structurally unique and dominant. Further details on these postulates and the six theorems derived from them can be found in Palak and Nguyen (2021)~\cite{bib16}.

\section*{Axiomatic Assessment of Centralization Measures}

We use the six postulates of Palak and Nguyen (2021)~\cite{bib16} as axioms to assess the suitability of each of the 11 centralization measures in our work. Table~\ref{tab:axioms} summarizes the results of these assessments. Proof ideas and counterexamples for our axiomatic assessments are provided in a narrative form as follows. 

\begin{table}[!ht]
\begin{adjustwidth}{-2.25in}{0in} % shift table left to center on full page
\centering
\caption{\textbf{Compliance of centralization measures with the six axioms proposed by Palak and Nguyen (2021) for evaluating network centralization.} A check mark (\ding{51}) indicates that the measure satisfies the corresponding axiom, whereas a cross (\ding{55}) indicates a violation. The final column reports the total number of satisfied axioms (maximum = 6). Abbreviations: ABH, Assortativity-Based Hubness; ECD, Eigenvector Centrality Dispersion; NBC, Normalized Betweenness Centralization; NCC, Normalized Closeness Centralization; NDC, Normalized Degree Centralization; NDE, Normalized Degree Entropy; NDV, Normalized Degree Variance; NGC, Normalized Gini Coefficient; NHD, Normalized Hub Dominance; NHT, Normalized Hub Formation Tendency; NNC, Normalized Natural Connectivity.}
\begin{tabular}{|l|c|c|c|c|c|c|c|}
\hline
\textbf{Measure} & \textbf{P1} & \textbf{P2} & \textbf{P3} & \textbf{P4} & \textbf{P5} & \textbf{P6} & \textbf{\# Satisfied} \\ \hline
ABH & \ding{51} & \ding{51} & \ding{51} & \ding{55} & \ding{55} & \ding{55} & 3 \\ \hline
ECD & \ding{51} & \ding{55} & \ding{51} & \ding{55} & \ding{55} & \ding{55} & 2 \\ \hline
NBC & \ding{51} & \ding{51} & \ding{51} & \ding{51} & \ding{51} & \ding{55} & 5 \\ \hline
NCC & \ding{51} & \ding{51} & \ding{51} & \ding{51} & \ding{51} & \ding{55} & 5 \\ \hline
NDC & \ding{51} & \ding{51} & \ding{51} & \ding{51} & \ding{51} & \ding{55} & 5 \\ \hline
NDE & \ding{51} & \ding{55} & \ding{51} & \ding{51} & \ding{55} & \ding{55} & 3 \\ \hline
NDV & \ding{51} & \ding{55} & \ding{51} & \ding{51} & \ding{55} & \ding{55} & 3 \\ \hline
NGC & \ding{51} & \ding{55} & \ding{51} & \ding{55} & \ding{55} & \ding{55} & 2 \\ \hline
NHD & \ding{55} & \ding{51} & \ding{51} & \ding{51} & \ding{51} & \ding{51} & 5 \\ \hline
NHT & \ding{55} & \ding{51} & \ding{51} & \ding{51} & \ding{55} & \ding{55} & 3 \\ \hline
NNC & \ding{51} & \ding{55} & \ding{51} & \ding{51} & \ding{51} & \ding{55} & 4 \\ \hline
\end{tabular}
\label{tab:axioms}
\end{adjustwidth}
\end{table}

ABH satisfies P1 because it behaves as expected for graphs related to P1. It satisfies P2 because the degree assortativity of a star graph is always $-1$. Since degree sequences and edge connections are preserved under isomorphism, ABH satisfies P3. It violates P4 because of counterexamples, including a graph with four nodes and two edges incident on the same node, which yields $ABH = 1$ despite having no saturated node. The graph with edge list $[(0,1), (0,2), (0,3), (0,4), (2,3), (3,4)]$ shows that ABH violates P5. The five-node graph with two edges incident on the same node shows that ABH violates P6.

ECD satisfies P1 (returns 0 for all graphs considered in P1) but violates P2 because star graphs do not realize the maximum standard deviation of eigenvector centrality. Since eigenvector centralities are preserved under isomorphism, ECD satisfies P3. It violates P4 because it can take value 1 in graphs without saturated nodes. ECD also violates P5, as demonstrated by the graph with the edge list $[(0,3), (0,4), (0,2), (1,3), (1,2), (1,4), (2,3), (3,4)]$. It violates P6 as shown by the 5-node graph with the edge list $[(0,1), (1,2), (2,3)]$.

NBC is claimed by Palak and Nguyen (2021)~\cite{bib16} to satisfy all six postulates. However, there is one error in that claim: the graph with edge list $[(0,1), (0,3), (0,4), (0,5), (1,2)]$ serves as a counterexample with respect to P6. For this graph, $NBC = 0.82$. Saturating node 3 reduces $NBC$ to 0.36, thereby violating P6.

NCC satisfies P1 (returns 0 for all P1 cases) and P2 (returns 1 for star graphs). Since closeness centralities are preserved under isomorphism, it also satisfies P3. Moreover, it satisfies P4 because in graphs without saturated nodes, the numerator is strictly smaller than the denominator. NCC satisfies P5, as saturating a node monotonically decreases its value. However, the 5-node graph with edge list $[(0,1), (0,2), (0,3)]$ shows that NCC violates P6.

NDC is claimed in Theorem 5 of Palak and Nguyen (2021)~\cite{bib16} to satisfy all six postulates. However, there is one error in that claim: the graph with edge list $[(0,1), (0,3), (0,4), (0,5), (1,2)]$ has $NDC = 0.7$. Saturating node 3 yields new edges $[(1,3), (2,3), (3,4), (3,5)]$ and reduces $NDC$ to 0.6, showing that it violates P6.

NDE satisfies P1 but violates P2, returning particularly low values for large star graphs. It satisfies P3 because degree sequences are preserved under isomorphism. It satisfies P4 because the theoretical maximum of Shannon entropy cannot be reached in simple graphs without self-loops. NDE violates P5, as shown by the graph with edge list $[(0,3), (0,2), (1,3), (1,4), (2,3), (3,4)]$. The 5-node graph that has one edge also shows that NDE violates P6.

NDV satisfies P1 but violates P2 because its normalization does not depend on star graphs. It satisfies P3 due to invariance over isomorphism. It also satisfies P4 because the maximum degree variance occurs in star graphs for $n < 7$ and in two-hub graphs for $n \ge 7$, both with at least one saturated node. NDV violates P5 as shown by the 7-node, 7-edge graph that is made up of a star structure with one extra edge. It violates P6, as shown by the graph with edge list $[(0,1), (0,2), (1,3), (1,4)]$.

NGC satisfies P1 but violates P2, taking value 0.5 for star graphs. It satisfies P3, as Gini coefficients are invariant under isomorphism. It violates P4, as it reaches 1 for one-edge graphs with isolated nodes. It also violates P5 as shown by the graph whose degree sequence is $\{3,3,3,3,4\}$. The 5-node graph that has one edge shows that NGC also violates P6.

NHD violates P1b, returning value 1 for complete graphs. It satisfies P2. The isomorphism invariance makes it satisfy P3. It satisfies P4, because graphs without saturated nodes always yield a numerator smaller than the denominator. It satisfies P5 because it returns value 1 for any graph with at least one saturated node. For the same reason, it also satisfies P6.

NHT violates P1b because it does not return 0 for complete graphs. It satisfies P2 and P3. It satisfies P4 because the only graphs where it returns 1 are star graphs and 3-node cycles of saturated nodes. It violates P5 as shown by the graph whose edge list is $[(0,1), (0,3), (0,2), (1,2), (1,3), (1,4), (2,4), (3,4)]$. The 5-node graph that has one edge shows that NHT also violates P6.

NNC satisfies P1 but violates P2, because the natural connectivity of star graphs is not exactly zero. NNC satisfies P3 because natural connectivity is isomorphism invariant. It satisfies P4 because it returns values below 1 when a saturated node is not present. It satisfies P5 because saturating a node increases natural connectivity and therefore decreases NNC. For the same reason, it violates P6.

\section*{Numerical Assessment of Centralization Measures}

To illustrate and compare the behavior of centralization measures under different network structures as $n$ grows, we analyze synthetic networks for which a desirable behavior is expected. These networks include star graphs, ring graphs, and complete graphs, along with their perturbed variants (same graph but with one edge rewired or removed). Figure~1 presents the values of the 11 centralization measures for one network type in each panel as $n$ grows. Table~\ref{tab:numerical} summarizes whether the measures returns values consistent with the expected behavior.

\begin{table}[!ht]
\begin{adjustwidth}{-2.25in}{0in} % shift table left to center on full page
\centering
\caption{\textbf{Performance of centralization measures across canonical graph topologies and their perturbed variants}. Measures were evaluated on star, ring, and complete graphs, as well as perturbed versions with a single edge rewired or removed. A check mark (\ding{51}) indicates expected behavior for the given topology, whereas a cross (\ding{55}) indicates deviation from expectations. The final column reports the number of topologies for which each measure performed as expected (maximum = 6). Abbreviations: ABH, Assortativity-Based Hubness; ECD, Eigenvector Centrality Dispersion; NBC, Normalized Betweenness Centralization; NCC, Normalized Closeness Centralization; NDC, Normalized Degree Centralization; NDE, Normalized Degree Entropy; NDV, Normalized Degree Variance; NGC, Normalized Gini Coefficient; NHD, Normalized Hub Dominance; NHT, Normalized Hub Formation Tendency; NNC, Normalized Natural Connectivity.}

\begin{tabular}{|l|c|c|c|c|c|c|c|}
\hline
\textbf{Measure} & \textbf{Star} & \textbf{Ring} & \textbf{Complete} & \textbf{Perturbed Star} & \textbf{Perturbed Ring} & \textbf{Perturbed Complete} & \textbf{\# Passed} \\ \hline
ABH & \ding{51} & \ding{51} & \ding{51} & \ding{51} & \ding{55} & \ding{55} & 4 \\ \hline
ECD & \ding{55} & \ding{51} & \ding{51} & \ding{55} & \ding{55} & \ding{51} & 3 \\ \hline
NBC & \ding{51} & \ding{51} & \ding{51} & \ding{51} & \ding{51} & \ding{51} & 6 \\ \hline
NCC & \ding{51} & \ding{51} & \ding{51} & \ding{51} & \ding{51} & \ding{51} & 6 \\ \hline
NDC & \ding{51} & \ding{51} & \ding{51} & \ding{51} & \ding{51} & \ding{51} & 6 \\ \hline
NDE & \ding{55} & \ding{51} & \ding{51} & \ding{55} & \ding{51} & \ding{51} & 4 \\ \hline
NDV & \ding{55} & \ding{51} & \ding{51} & \ding{55} & \ding{51} & \ding{51} & 4 \\ \hline
NGC & \ding{55} & \ding{51} & \ding{51} & \ding{55} & \ding{51} & \ding{51} & 4 \\ \hline
NHD & \ding{51} & \ding{55} & \ding{55} & \ding{51} & \ding{51} & \ding{55} & 3 \\ \hline
NHT & \ding{51} & \ding{55} & \ding{55} & \ding{51} & \ding{51} & \ding{51} & 4 \\ \hline
NNC & \ding{55} & \ding{55} & \ding{51} & \ding{51} & \ding{55} & \ding{51} & 3 \\ \hline
\end{tabular}
\label{tab:numerical}
\end{adjustwidth}
\end{table}

The star graph is expected to get values reflecting maximal centralization, consistent with the common assumptions in the literature~\cite{bib1,bib16,bib28}. In the top left panel of Figure~1, the six measures, ABH, NBC, NCC, NDC, NHD, and NHT, correctly return the value of 1, reflecting maximal centralization regardless of the network order. ECD and NNC also return high values as network order increases but do not reach 1. Conversely, NDV and NGC produce values distinctly below 1. NDE shows values substantially below 1 for small network sizes and unexpectedly trending toward zero as network order increases, indicating its crucial weakness in quantifying maximal centralization.

In the top center panel of Figure~1, we move on to the ring graph where all nodes have identical degrees and centralities. In this case, we expect a centralization value of zero due to the lack of a distinct or dominant node. This expectation is confirmed by eight measures, ABH, ECD, NBC, NCC, NDC, NDE, NDV, and NGC, all returning zero for all ring networks. However, NHD and NHT show a gradual decay toward zero as network order increases. NNC behaves unintuitively, failing to return zero and instead increasing to large values as network order grows, indicating its unsuitability from this perspective.

The complete graph represents maximal connectivity among structurally identical nodes. Therefore, centralization measures are expected to return a value of 0 for complete graphs in the top right panel of Figure~1. This is indeed the case for nine measures, ABH, ECD, NBC, NCC, NDC, NDE, NDV, NGC, and NNC, across all network orders, reaffirming their validity under axiom P1b. NHT exhibits values with a decreasing trend toward zero as complete graphs grow. NHD remains fixed at 1, highlighting a critical limitation in its ability to reflect an intuitive value for complete graphs.

Rewiring a single edge in the star graph slightly disrupts the hub configuration, but this alteration is expected to become increasingly negligible as the network order grows. The seven measures, ABH, NBC, NCC, NDC, NHD, NHT, and NNC, reflect this expected trend in the bottom left panel of Figure~1. In contrast, NDV and NGC consistently remain far from 1 and are largely insensitive to the rewiring and its interaction with network order. NDE incorrectly shows a steady decline with increasing network order, and ECD shows an increasing trend that does not seem to converge to 1, indicating a poor response to such a perturbation.

In the perturbed ring graph, rewiring one edge introduces a slight increase in centralization, which is expected to diminish as the network order increases. All measures except ABH, ECD, and NNC capture these subtle changes in the bottom center panel of Figure~1. They correctly reflect the existence of nodes that are slightly more hub-like than others (against an otherwise random-regular structure without any distinct nodes). ABH, ECD, and NNC incorrectly exhibit a constant non-zero value or an increasing trend toward 1, revealing their limitations in accurately representing intuitive values for centralization in the rewired ring graphs.

Removing a single edge from the complete graph introduces minimal asymmetry, as the graph remains nearly fully connected. The bottom right panel of Figure~1 shows that despite the subtlety of this perturbation, measures such as ABH and NHD fail to approach zero. They are insensitive to slight deviations from complete connectivity. Other measures, however, show an appropriate decline toward zero, indicating a suitable capacity to reflect the centralization of such perturbed complete graphs.

In summary, across all the graph types and perturbations considered, NBC, NCC, and NDC demonstrate the most robust comparative performance, consistently reaching expected boundary values and responding smoothly and intuitively to structural changes. Other measures show expected behavior in some aspects but behave inconsistently in others. Table~\ref{tab:numerical} summarizes how each measure behaves for the six network types as $n$ increases.

\begin{figure}[!ht]
\centering
\includegraphics[width=\textwidth]{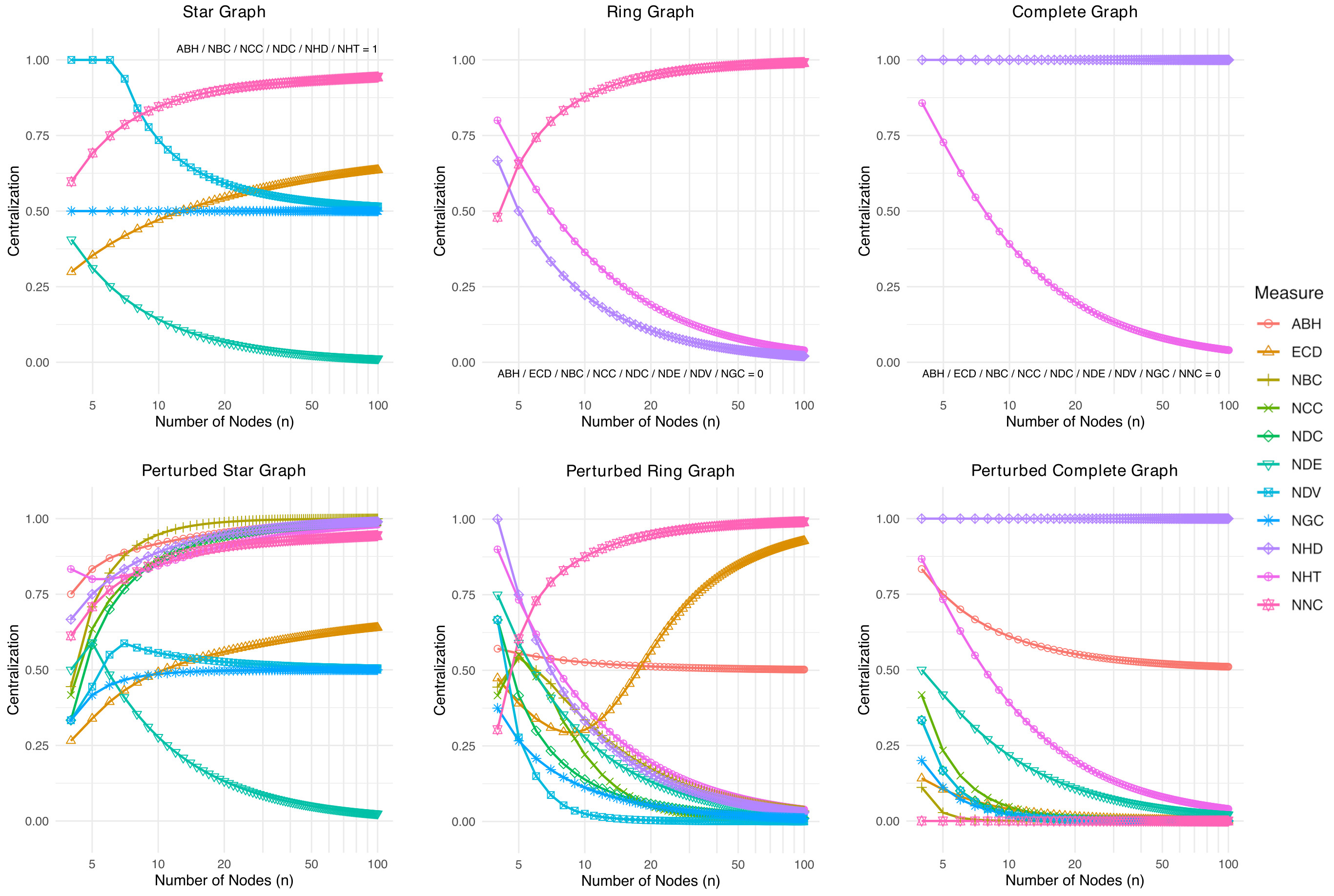} % Replace with your actual file name
\caption{\textbf{Centralization values computed using various measures across different network sizes for three graph configurations: star, ring, and complete graphs (top row), and their perturbed variants with a single edge rewired or removed (bottom row).} Each plot shows how each measure responds to changes in network order. Measures that yield constant values of 0 or 1 are not shown and are instead annotated in subplots. Colors and point styles indicate different centralization measures. Abbreviations: ABH, Assortativity-Based Hubness; ECD, Eigenvector Centrality Dispersion; NBC, Normalized Betweenness Centralization; NCC, Normalized Closeness Centralization; NDC, Normalized Degree Centralization; NDE, Normalized Degree Entropy; NDV, Normalized Degree Variance; NGC, Normalized Gini Coefficient; NHD, Normalized Hub Dominance; NHT, Normalized Hub Formation Tendency; NNC, Normalized Natural Connectivity.}

\label{fig:centralization}
\end{figure}

\section*{Overall Evaluation of Centralization Measures}

To combine the axiomatic and numerical assessments, we define a simple additive overall performance score as a weighted sum of the Axiomatic Score $S_A$ (number of axioms satisfied, maximum = 6) and the Numerical Score $S_N$ (number of numerical simulations passed, maximum = 6):
\begin{equation}
\mathrm{Total\ Score} = w_A S_A + w_N S_N
\end{equation}
where $w_A$ and $w_N$ are the weights assigned to the axiomatic score and numerical score respectively. The two weights add up to 1 $w_A + w_N = 1$. Different weightings can be chosen based on the researcher’s priorities, possibly putting different emphasis on theoretical validity in foundational work or on numerical robustness in applied analyses. 

In our assessment study, we set $w_A = w_N = 0.5$, assigning equal importance to both criteria. Under this weighting, the measures NBC, NCC, and NDC emerge as the strongest overall performers (aggregate score of 5.5 each), achieving near-perfect performance and demonstrating both strong axiomatic compliance and robust numerical behavior.

NHD (scoring 4) performs well on the axiomatic criteria but shows certain weaknesses in numerical behavior, particularly in regular graphs. ABH, NDE, NDV, NHT, and NNC (each scoring 3.5), as well as NGC (scoring 3), display some elements of intuitiveness, showing partial numerical consistency but weak theoretical alignment with the axioms. At the lower end, ECD (scoring 2.5) performs poorly in both domains, indicating substantial limitations in capturing centralization consistently.

This simple evaluation framework provides a flexible and transparent decision-making tool for selecting centralization measures that align with the priorities of a given study, whether focused on theoretical validity, numerical robustness, or a balanced combination of both.

\section*{Application to Real-World Networks}

Now that the existing measures are refined down to three measures, we can demonstrate their practical utility. We computed NBC, NCC, and NDC for a diverse set of real-world networks spanning different domains (Table~3). Given that NBC and NCC rely on path-based distances, they cannot be used in disconnected graphs. Therefore, we use the largest connected component of multi-component real networks in this section to demonstrate the utility of all three measures NBC, NCC, and NDC.

These networks vary substantially in size, density, and structural characteristics, providing a robust demonstration of each measure’s behavior beyond synthetic benchmarks. For networks \textit{Condensed Matter Collaboration} and \textit{Yeast Protein Interactions}, all three measures yield relatively low values ($NBC \le 0.158$, $NCC \le 0.217$, $NDC \le 0.081$), reflecting the absence of a single dominant hub and a relatively even distribution of connectivity. These networks are relatively large and sparse. In contrast, networks with more hierarchical structures, such as the \textit{Food Web (Michigan)} and \textit{Zachary’s Karate Club}, exhibit substantially higher centralization: NBC (0.196/0.405), NCC (0.701/0.298), and NDC (0.642/0.399). These networks are smaller and more dense. The values indicates the presence of central nodes and their relative extent of dominance. 

The relatively lower NBC value in the \textit{Food Web (Michigan)} suggests that different centralization measures are not necessarily associated and instead capture distinct properties. Interestingly, these three measures often diverge in their sensitivity to network structures, highlighting that they capture complementary aspects of centralization. For example, the \textit{Facebook Ego Network} shows a markedly higher NBC value (0.480) compared to NDC (0.248), with NCC lying in between (0.367). This suggests that path-based centralization in this network dominate over degree-based centralization, while closeness captures an intermediate notion of reachability in this specific social media network. Conversely, the \textit{Brain Network (HCP)} shows moderate NBC (0.062) but relatively high NCC (0.288) and NDC (0.291), consistent with a densely connected hub structure in functional brain connectivity~\cite{bib14,bib15}.

These results highlight that while NBC, NCC, and NDC often agree in identifying low- or high-centralization networks, their differing definitions capture potentionally complementary aspects of centralization. NBC reflects path-based control, NCC emphasizes reachability and distance efficiency, and NDC quantifies degree-based hub concentration. When applied together, they can reflect a richer and more reliable picture of network centralization from different aspects compared to what any single measure can reflect alone.

\begin{table}[!ht]
\begin{adjustwidth}{-2.25in}{0in} % use full page width (PLOS ONE style)
\centering
\caption{\textbf{Normalized Betweenness Centralization (NBC), Normalized Closeness Centralization (NCC), and Normalized Degree Centralization (NDC) for diverse real-world networks.} 
For each network, the category, description, number of nodes ($n$), and number of links ($m$) are provided. 
All values are computed on the largest connected component of each network.}
\renewcommand{\arraystretch}{1.05}
\setlength{\tabcolsep}{4pt}
\begin{tabular}{|p{3.0cm}|p{2.8cm}|p{5.2cm}|p{1.2cm}|p{1.4cm}|c|c|c|}
\hline
\textbf{Network} & 
\textbf{Category} & 
\textbf{Description} & 
\centering \textbf{\# Nodes} \\ \textbf{($n$)} & 
\centering \textbf{\# Links} \\ \textbf{($m$)} & 
\textbf{NBC} & 
\textbf{NCC} & 
\textbf{NDC} \\ \hline
Brain Network (HCP)~\cite{bib35} & Neuroscience / Biological & Functional coactivations between brain regions & 45 & 242 & 0.062 & 0.288 & 0.291 \\ \hline
Condensed Matter Collaboration~\cite{bib36} & Scientific Collaboration & Co-authorship network among physicists publishing on condensed matter & 21363 & 182628 & 0.088 & 0.217 & 0.025 \\ \hline
Enron Email Network~\cite{bib37,bib38} & Organizational / Social & Email communications between employees of Enron corporation & 36692 & 361622 & 0.077 & 0.269 & 0.081 \\ \hline
Facebook Ego Network~\cite{bib39} & Social Media & Social connections on Facebook (ego-centered) & 4039 & 88234 & 0.480 & 0.367 & 0.248 \\ \hline
Food Web (Michigan)~\cite{bib40} & Ecological / Biological & Predator–prey relationships in the Michigan ecosystem (converted to undirected) & 39 & 212 & 0.196 & 0.701 & 0.642 \\ \hline
Yeast Protein Interactions~\cite{bib41} & Biological & Interactions between proteins in yeast, forming a PPI network & 2375 & 11693 & 0.158 & 0.206 & 0.046 \\ \hline
Zachary’s Karate Club~\cite{bib42} & Social & Friendship network among 34 members of a karate club & 34 & 78 & 0.405 & 0.298 & 0.399 \\ \hline
\end{tabular}
\label{tab:realworld}
\end{adjustwidth}
\end{table}

We also examined a temporal friendship network among high-school students, observed across four survey waves during one academic year~\cite{bib43}. The values of NBC, NCC, and NDC fluctuated over time, revealing dynamic shifts in centralization. In Month~1, NBC (0.22), NCC (0.25), and NDC (0.22) were low, reflecting a relatively decentralized and fragmented structure without clear hubs. By Month~2, NBC (0.52), NCC (0.85), and NDC (0.75) peaked, indicating the emergence of a dominant hub both in terms of control over network paths, geodesic proximity, and direct connectivity. In Month~3, NBC dropped sharply (0.10), while both NCC (0.66) and NDC (0.54) remained moderately high, suggesting a transition toward well-connected nodes that preserved degree- and closeness-based centralization without a single dominant path bottleneck. Finally, in Month~4, NBC stabilized at an intermediate level (0.36), NCC rose again (0.82), and NDC increased (0.71), highlighting the persistence of degree- and closeness-based hub concentration alongside more distributed path-based centralization. Taken together, these results show that tracking the three measures reveals a multi-faceted profile of centralization for each snapshot of a network evolving over time.

\begin{figure}[!ht]
\centering
\includegraphics[width=\textwidth]{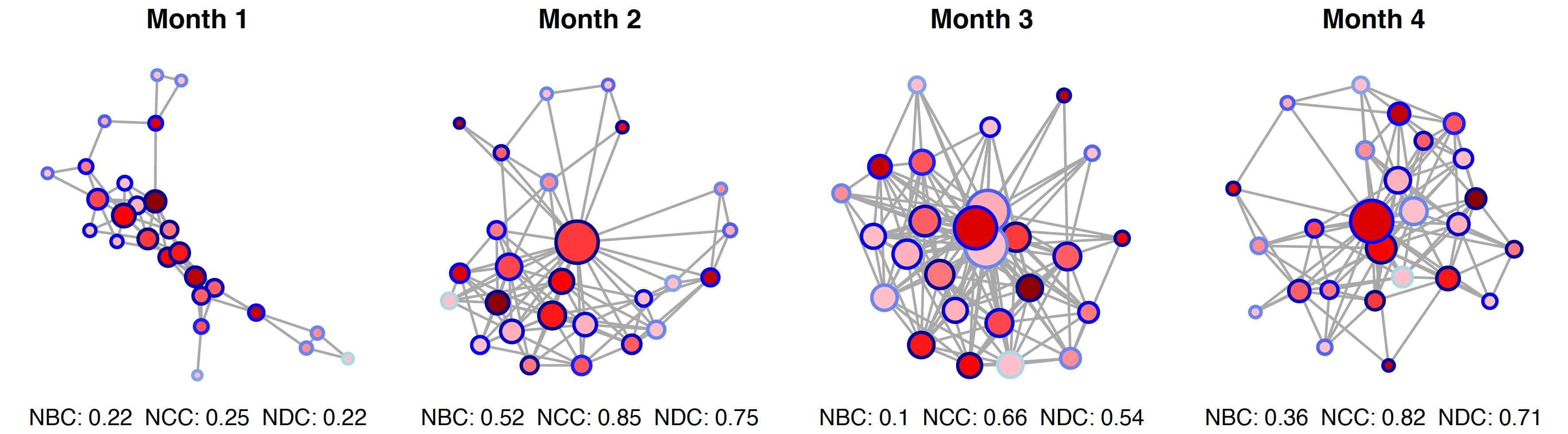} % replace with your actual file name
\caption{\textbf{Temporal snapshots of a high-school friendship network across four survey waves.} Nodes represent individuals, and links denote reciprocated friendship ties. Node size is proportional to degree, and node color intensity reflects betweenness centrality (darker red = higher values), while node border color reflects closeness centrality (darker blue = higher values). Normalized Betweenness Centralization (NBC), Normalized Closeness Centralization (NCC), and Normalized Degree Centralization (NDC) values are reported below each panel.}
\label{fig:highschool}
\end{figure}

All simulations, network analyses, and visualizations were conducted using R, utilizing several packages including \texttt{igraph}~\cite{bib44} for network modeling and \texttt{ggplot2}~\cite{bib45} for visualization. Code to replicate the results and extend the research is available at: \url{https://github.com/majidsaberi/Centralization}.

\section*{Discussion}

\subsection*{Summary of Key Findings}

This study reviewed and normalized 11 centralization measures from existing approaches and provided an evaluation of them using axioms and numerical simulations. The evaluation combined axioms based on the six postulates of Palak and Nguyen (2021)~\cite{bib16}, with numerical assessments on synthetic networks exhibiting extreme centralization behaviors. This dual evaluation revealed that Normalized Betweenness Centralization (NBC), Normalized Closeness Centralization (NCC), and Normalized Degree Centralization (NDC) consistently behaved better than other measures. These three metrics not only satisfied most axioms but also demonstrated stable behavior consistent with the expected behavior across different network types.

Application of these best-performing measures to a diverse set of real-world networks revealed new insights into the complementary nature of the three measures. In several cases, NBC, NCC, and NDC aligned closely in identifying highly centralized systems but from three distinct perspectives. However, in networks such as the \textit{Facebook Ego Network} and \textit{Brain Network}, the values of the three measures diverged substantially, reflecting their distinct underlying definitions. NBC highlights path-based dominance, NCC captures closeness-based accessibility, and NDC emphasizes degree-based hub concentration. Furthermore, analysis of a temporal friendship network showed that NBC, NCC, and NDC capture different dynamic expressions of centralization over time.

These results confirm that no single measure fully captures the multi-faceted notion of centralization that appears from a joint usage of selective suitable measures. While our results point to three measures, we argue that the narrower choice of measures can be guided by the definition of centralization and dominance (path-based, closeness-based, or degree-based) that is most relevant to the research question.

\subsection*{Methodological Implications}

Our results highlight the benefits of assessing centralization measures using a framework that integrates axiomatic validation and numerical performance testing. The axiomatic framework of Palak and Nguyen (2021)~\cite{bib16} provides a theoretical baseline, ensuring that a measure behaves consistently under well-defined structural conditions such as maximal centralization, minimal centralization, and changes in centralization. Notably, some measures satisfied several axioms yet displayed counterintuitive trends on synthetic networks. Therefore, axiomatic compliance can be enhanced by numerical simulations.

Our assessment enables researchers to identify metrics that are both conceptually valid and operationally reliable, while also allowing flexibility to diverge from our specific choices to better satisfy the constraints of their study. Our approach is fairly general and can be adapted to other network contexts, including weighted, directed, or temporal networks, where similar concerns about measure validity and stability are likely to arise.

\subsection*{Choosing Between NBC, NCC, and NDC Based on Network Context}

Our findings indicate that among the 11 existing measures, NBC, NCC, and NDC are arguably more robust choices for quantifying centralization in undirected unweighted networks, and they capture distinct aspects of centralization.

NBC is particularly suitable for studying networked systems where control over network paths or flow bottlenecks is of primary importance. These systems may include transport networks, communication infrastructures, or networks of organizational information flow. NCC, by contrast, is especially relevant in contexts where accessibility and efficiency of reach are central concerns, such as emergency response networks, social contact networks, or information dissemination systems, where certain nodes minimize average distances to others. NDC is most appropriate for contexts where the concentration of direct connectivity (degree) is a defining factor, such as biological hub proteins, brain network hubs, or models of critical junctions in infrastructure networks.

The interpretation of centralization in empirical settings benefits from the complementary deployment of multiple well-behaved measures (possibly NBC, NCC, and NDC). This is evident in the \textit{Facebook Ego Network}, where NBC yielded a markedly higher score than NDC. This suggests that while no single node dominated in degree, the structure allows several nodes to exert strong influence over communication flow. Meanwhile, NCC provided an additional perspective on how central nodes reduced distances and improved accessibility across the network. Conversely, in the \textit{Brain Network (HCP)}, NCC and NDC were substantially higher than NBC. This reflected dense connectivity around key hub regions and the efficiency of those hubs in reaching other regions, without equivalent dominance in terms of path-based control. These contrasts illustrate how NBC, NCC, and NDC can together provide complementary insights into the subtle, nuanced, and multi-faceted notion of network centralization.

Beyond static networks, the complementary use of NBC, NCC, and NDC also extends to temporal settings, where they can highlight changes to a dynamic profile of centralization over time.

\subsection*{Extending Centralization to Weighted, Directed, and Sparse Networks}

Many real-world networks are inherently weighted (e.g., transportation intensity, communication frequency, and synaptic strength) or directed (e.g., citation network, neuronal signaling, and trade relations). Therefore, extending centralization measures beyond undirected and unweighted graphs is an essential next step toward a more general and realistic framework. Although substantial research has focused on extending node-level centrality metrics to weighted and directed cases~\cite{bib7}, global centralization measures have received far less systematic attention or formal generalization. Extending our axiomatic–numerical framework to weighted and directed networks presents both conceptual and computational challenges: 

First, directionality introduces asymmetry, where incoming and outgoing influences can differ between nodes. In this context, all three measures require distinct adjustments. NDC can be divided into in-degree and out-degree centralization, each normalized against the theoretical maximum for a directed star graph (one node receiving from or sending to all others). For NBC, directionality alters shortest-path dependencies. So, paths must be computed with directional constraints, capturing asymmetries in information or flow control. Likewise, NCC in directed networks must account for reachability asymmetry, often necessitating separate in- and out-closeness variants to capture accessibility versus broadcasting efficiency. 

Second, weights complicate the notion of maximal centralization because dominance can now arise either through the number of links or through their strength. Extending NDC involves computing degree strength (sum of edge weights) rather than simple counts, normalized against the strongest possible star configuration. For NBC, shortest paths should be computed using weighted distances (e.g., inverse weights) while maintaining normalization relative to a weighted-star reference. NCC can similarly treat weights as inverse distances, thus reflecting the efficiency of strong versus weak ties. 

Third, a fundamental theoretical issue concerns the definition of a reference configuration for normalization. In unweighted networks, the star graph uniquely represents maximal centralization. In weighted or directed systems, however, multiple “weighted-star” or “asymmetric-star” structures could yield the same level of dominance depending on how weights are distributed. This ambiguity highlights the need to redefine reference models that preserve both total strength and connectivity constraints, ensuring that normalization remains scale-consistent.

The same axiomatic approach deployed in this study can guide generalizations to weighted and directed networks. Each extended measure should be examined for compliance with the six postulates under asymmetric and weighted conditions. This will reveal whether measures such as NBC, NCC, and NDC retain desirable behaviors when the underlying assumptions of symmetry and binary connectivity are relaxed.

Beyond directionality and weighting, another dimension of generalizability concerns sensitivity in sparse networks. Many empirical systems, especially social, biological, and infrastructural networks, are sparsely connected and subject to small perturbations. Sparse networks tend to amplify centralization effects: the removal or addition of a single hub connection can disproportionately shift global measures. Among the three selected metrics, NDC is generally more sensitive to small link changes because it depends directly on degree distribution, while NBC exhibits moderate sensitivity due to path recalculation in shortest routes. NCC tends to respond more smoothly, as changes in a few links have limited influence on global accessibility unless they occur near key bottlenecks. Quantifying such perturbation sensitivity could be formalized by measuring stability under random edge perturbations, temporal rewiring, or bootstrapped network re-sampling.

A further computational and interpretive challenge arises in multi-component networks, which are common in sparse real-world systems. For such networks, path-based centralization measures like NCC and NBC cannot be directly computed for the entire graph because path lengths between disconnected components are undefined. This issue can distort normalization and lead to misleadingly low or undefined centralization scores. Future extensions of the framework could explore component-weighted aggregation schemes, penalized path-length approximations, or effective distance formulations that preserve interpretability while enabling valid comparison across disconnected networks. 

\subsection*{Limitations and Future Directions}

While this study provided a comparative evaluation of 11 normalized centralization measures, several limitations should be acknowledged. First, our analyses focused exclusively on undirected and unweighted and connected networks. %Many real-world systems include directionality (e.g., citation networks, neural connectivity) or weighted relationships (e.g., traffic volumes, communication frequencies), which can substantially influence centralization patterns. 
The present framework requires major adjustments before being applicable to weighted or directed networks as discussed earlier. %Our preliminary exploration of temporal networks suggests that centralization in networked systems may evolve over time, underscoring the benefits of extending such evaluations beyond static assessments of a single snapshot of a system, where applicable.

Second, our numerical assessment was inherently binary, each measure was considered to either satisfy or fail a given numerical assessment. In practice, violations may vary in magnitude, and a more nuanced scoring system could provide finer-grained insight into the partial compliance that a measure demonstrates for a given graph type.

Third, although NBC and NCC were identified alongside NDC as relatively strong measures, both rely on distance-based properties of networks. This makes their interpretation problematic in disconnected or multi-component graphs, since shortest paths between components are undefined. In this study, we addressed this challenge by computing NBC and NCC only on the largest connected component of each real-world network. While this ensures comparability and interpretability, it may exclude hub-like structures that exist in smaller disconnected components. Future work should therefore examine the robustness of these measures under broader conditions, including sparse or fragmented networks, or by exploring alternative path-based formulations that account for multi-component networks.

Fourth, as discussed, sparse and near-sparse networks amplify centralization effects: small perturbations, such as the removal or addition of a single hub link, can substantially alter global measures. Systematic sensitivity analyses, for instance, through random edge perturbations or temporal snapshots, could help quantify each measure’s stability under minimal structural change.

Finally, although this study focused on theoretical validity and empirical stability, computational efficiency was not the primary objective in our evaluation. For extremely large-scale networks or time-constrained analyses, algorithmic efficiency may become an important consideration in narrowing down the selection of measures among the well-behaved measures that we have identified.

\subsection*{Conclusion}

Our results show that NBC, NCC, and NDC emerged as the three most robust metrics for quantifying network centralization among 11 existing approaches. NBC effectively captures path-based dominance, highlighting the influence of nodes that control communication flow. NCC captures accessibility-based dominance, reflecting the ability of certain nodes to minimize distances and improve efficiency of reach across the network. NDC quantifies degree-based hub concentration and the dominance of nodes with exceptional connectivity. When applied jointly, these three measures provide complementary insights into centralization compared to what any single measure can capture alone.

Our dual evaluation framework benefited from combining theoretical soundness with empirical robustness in metric selection. This framework clarifies conceptual distinctions among measures and assists researchers in choosing appropriate centralization metrics. Our findings encourage a shift away from reliance on single, conventional measures toward evaluating centralization as a nuanced multi-faceted notion, ultimately improving the our understanding of a topological feature that emerges in networks across different fields.

%This work advances both the methodological foundation and the applied utility of centralization metrics. 

%\nolinenumbers

%\bibliographystyle{plos2025}
%\bibliography{References}

\end{document}